\RequirePackage{fix-cm}
\documentclass[twocolumn]{svjour3} 
\smartqed
\usepackage{color,graphicx,hyperref,amsmath}
\usepackage{mathtools}
\usepackage{graphicx}
\usepackage{float}
\usepackage{cite}

 \journalname{Eur. Phys. J. Spec. Top.}
\begin{document}

\title{Analysis of the second wave of COVID-19 in India based on SEIR model}
\author{M. Manoranjani, Shamik Gupta and V. K. Chandrasekar}

\author{R. Gopal, V. K. Chandrasekar and M. Lakshmanan}

\institute{R. Gopal and V. K. Chandrasekar \at
	Departmrnt of Physics, Centre for Nonlinear Science and Engineering, School of Electrical and Electronics Engineering, SASTRA Deemed University, Thanjavur 613 401, India\\
	%
\and M. Lakshmanan \at  Department of Nonlinear Dynamics, School of Physics, 
Bharathidasan University, Tiruchirappalli -620 014, Tamil Nadu, India.\\}

\maketitle

\begin{abstract}
\par India was under a grave threat from the second wave of the COVID-19 pandemic particularly in the beginning of May 2021. The situation appeared rather gloomy as the number of infected individuals/active cases had increased alarmingly during the months of May and June 2021 compared to the first wave peak. Indian government/state governments have been implementing  various control measures such as lock-downs, setting up new hospitals, and putting travel restrictions at various stages to lighten the virus spread from the initial outbreak of the pandemic. Recently, we have studied the susceptible-exposed-infectious-removed(SEIR) dynamic modelling of the epidemic evolution of COVID-19 in India with the help of appropriate parameters quantifying the various governmental actions and the intensity of individual reactions. Our analysis had predicted the scenario of the first wave quite well. In this present article, we extend our analysis to estimate and analyze the number of infected individuals  during the second wave of COVID-19 in India with the help of the above SEIR model.  Our findings show that the people's individual effort along with governmental actions such as implementations of curfews, accelerated vaccine strategy are the most important factors to control the pandemic in the present situation and in the future. 
\keywords{COVID-19 \and SEIR model \and  infected individuals \and  } 
\end{abstract}
\section{Introduction}

\par In recent times, mathematical models for epidemiological studies have proved to be helpful during the Corona-virus pandemic by offering snapshots of specific patterns such as virus spread, identifying the number of infected individuals, upcoming consequences and so on~\cite{cohen,rothan,lin,li,nld}. In order to track the transmission of the virus, one has to essentially rely on the available data and make simplified assumptions to reduce the complexity of the problem. Furthermore, one has to keep oneself up to date with all the data and observations in order to develop models that are as close to reality as possible. This would also help in the evaluation of various scenarios, and for providing information for proper support for health care decisions.

 The affected countries have put in a lot of effort to counter the consequences of COVID-19. The crisis, however, is far from over, with newly infected people being identified every day. It is critical to create adequate models to characterize and understand the  change in the trend of the pandemic in order to estimate when the pandemic will be under control and to evaluate the effectiveness of virus control efforts. 
 
 \begin{figure*}
\centering 
\includegraphics[width=1.2\columnwidth]{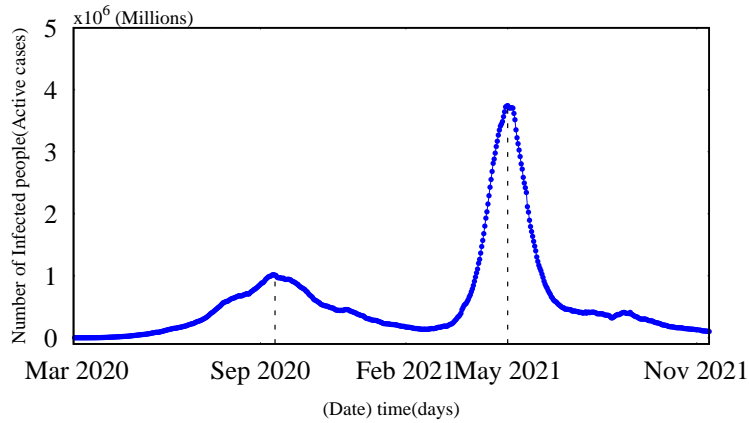}
\caption{The occurrence of the total number of active cases in the epidemic curve from March 2020 to November 2021 in India.}
\label{real}
\end{figure*}

\begin{figure}
\centering 
\includegraphics[width=1.0\columnwidth]{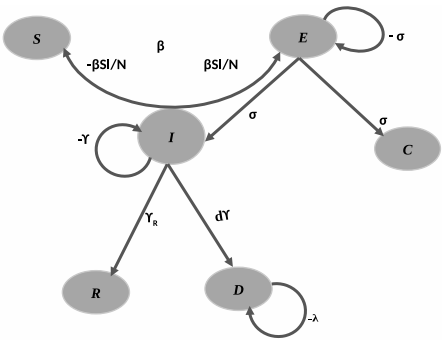}
\caption{Abstract model for COVID-19 dynamics based on SEIR framework.}
\label{c}
\end{figure}

\begin{figure*}
\centering 
\includegraphics[width=2.0\columnwidth]{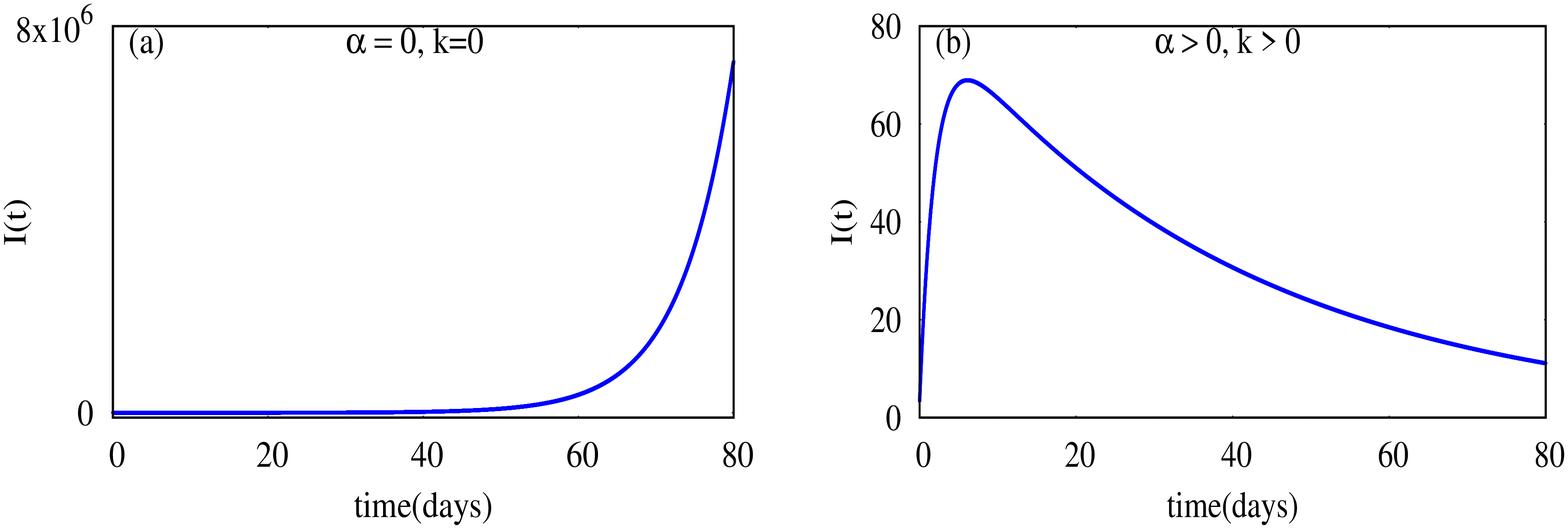}
\caption{Variation of number of the infected individuals for different values of $\alpha$ and $k$. }
\label{function}
\end{figure*}

  Recently, we have examined the commonly used susceptible-exposed-infected individuals-removed (SEIR) model~\cite{lin,li,he} to account for new uncertainties, and in particular, the number of  infected individuals in India which were estimated during the initial lock-down and unlock periods, starting from March, 25, 2020 to October, 31, 2020 and further up to December 27, 2020, with the help of the initial rate of COVID-19  transmission by considering the initially infected people in the country~\cite{gopal,gov}. The prediction of the infected individuals from our model has agreed with the actual data of the daily rate of infected individuals during the first wave fairly well~\cite{gopal,gov}. In order to predict and identify the evolution of COVID-19 virus spread, a series of research works have been recently proposed, and in particular the SIR and linear fractal based models have predicted the daily positive cases of the outbreaks of the second wave of COVID-19 in India and these are reported in Refs.~\cite{post,kavi}. Further, the dynamic evolution of the SEIR model along with various parameters, like incidence rate, transmission rate, test positivity rate, case fatality rate  and  intervention parameters were illustrated to describe the first and second waves of COVID-19 (See \cite{chen, ranjan,mus}).
  
  Figure \ref{real} shows the variation of active cases (number of individually infected people)  up to November, 2021. The data has been taken from the repository maintained by the Indian government official website and 'worldometers'~\cite{gov}. The graph shows that the occurrence of first and second waves of COVID - 19 was separated by almost 7-8 months. The peak of the active cases in the first wave had occurred in Sep 2020, and decreased around the middle of February 2021. The end of the first wave of COVID-19 is only due to the more public response along with effective government interventions~\cite{gopal}. Unfortunately, in the month of April 2021, the evolution of active cases has suddenly started to increase  and it shows more than double that of the first peak value by May 2021. The second wave in India appears to be considerably riskier than the first wave, but the situation is fast under control due to drastic measures such as lock-downs in various areas and increased vaccination campaigns.  The aim of this paper is to interpret and model the occurrence of the COVID-19 second wave, as well as to investigate the role of individual reactions in association with governmental actions, in order to raise public awareness about the COVID-19 and to highlight the importance of individual actions.

Therefore, in this present study, we again divided the common population into four compartments - susceptible, exposed, infectious and removed (which include both the cases of recovered and death numbers) with the help of appropriate differential equations to relate the parameters of the model to the population based on the initial framework of He {\it et al} which was proposed as a basic model that explained the 1918 influenza~\cite{he,savi} pandemic. Further, the specific aim of the proposed model is to capture the effects of the individual actions/responses (which may include personal hygiene, healthy habits, avoiding crowded places, wearing masks, washing hands frequently, taking vaccination), in addition to the governmental action (implementation of curfew and acceleration of vaccine strategy) both of which are represented by appropriate parameters.

  In this paper, we further consider the variability in transmission rate through the values of individual action rate and governmental action strength from an all India basis as well as in the various individual states (particularly Tamilnadu, Maharashtra, Kerala, and Karnataka as typical examples),  starting from Dec 2020.  In particular, we examine the number of infected individuals due to the occurrence of the second wave of COVID -19, and predict the possible peaks as well as validate the importance of intensity of individual reaction along with the governmental action control during the pandemic.  So far, during the pandemic period, several factors have had an impact on whether the number of infected individuals  is increasing or declining at particular locations. These factors include human behaviour, infection prevention policies, changes of the COVID-19 virus itself, effectiveness of vaccines over time. Our study shows that significant spread of the number of infected individuals can raise during the second wave of COVID -19 and that the control of the disease in a short period is possible mainly due to the intensity of individual reaction ($k$) along with governmental action strength $\alpha$.  Therefore, we feel that our study indicates that once people follow the COVID 19 precautions, such as getting vaccinated for the COVID 19, predicting physical distance,  hand-washing and mask-wearing, etc., control of the spread of COVID -19 becomes feasible.

  Based on the above objectives, the remaining part of the paper is organized as follows. In Section II, we describe the modified SEIR mathematical model and the analysis of its dynamics with help of intensity of individual reaction and governmental action strength. In Section III, we analyze the occurrence of the number of infected individuals and discuss the results of our analysis with help of reproduction numbers. Finally, we summarize our findings in section IV.
  
\section{Description of the model}  
  
The suggested  SEIR  model has been formulated as follows~\cite{lin,li,gopal,he,savi} based on the original work of He et al~\cite{he},

\begin{subequations}
\begin{eqnarray}
&&\dot{S}=-\beta(t)\frac{SI}{N},  \\
\label{eq1b}
&&\dot{E}=\beta(t)\frac{SI}{N}-\sigma~E ,  \\
\label{eq1c}
&&\dot{I}=\sigma E-\gamma I , \\
&&\dot{R}=\gamma_{R} I,  \\
&&\dot{D}=d\gamma I-\lambda D ,  \\
&&\dot{C}=\sigma E.  
\label{eq1}
\end{eqnarray}
\label{eq1}
\end{subequations}

In this model (\ref{eq1}), $S$ is the susceptible population, $E$ is the exposed population, $I$ is the currently infected population (excluding the recovered and death cases)  and $R$ is the population removed which includes both the cases of recovered and death numbers. Also, $N$ is the total number of population. Besides the above, the total population also contains two more classes: $D$ is a public perception of risk with respect to serious cases and deaths, and $C$ is the number of cumulative cases (both reported and not reported incidents)~\cite{lin}. Based on this, it is possible to establish an abstract model of the  COVID-19 dynamics presented in Fig. \ref{c}
which allows one to write the above equations (\ref{eq1}). 
Further, the following parameters are also considered in the governing equations: $\gamma$ is the mean infectious period, $\gamma_{R}$ is the delayed removal period, which denotes the relation between the removed population and the infected ones, $\sigma$ is the mean latent period, while $d$ denotes the proportion of severe cases and $\lambda$ is the mean duration of public reaction~\cite{li,lin,savi}.

In Eq.(\ref{eq1}), $\beta(t)$ denotes the transmission rate function which incorporates the impact of governmental action $(1-\alpha)$, and the individual action, which is denoted by the function $\left (1-\frac{D}{N} \right )^{k}$~\cite{lin,savi}.   Here, the parameter $k$ defines the intensity of individual reaction, which is measured on a scale of 0 to $10^5$ with a normal value of 1117.3 obtained from previous and recent epidemic and pandemic studies~\cite{he,lin}. These values should be incorporated into the different values of the transmission rate which is defined as~\cite{lin} 
\begin{align}
\beta(t)=\beta_{0}(1-\alpha) \left (1-\frac{D}{N}\right)^{k}. 
\label{eq2}
\end{align}

\begin{figure*}
	\centering
	\includegraphics[width=2.0\columnwidth]{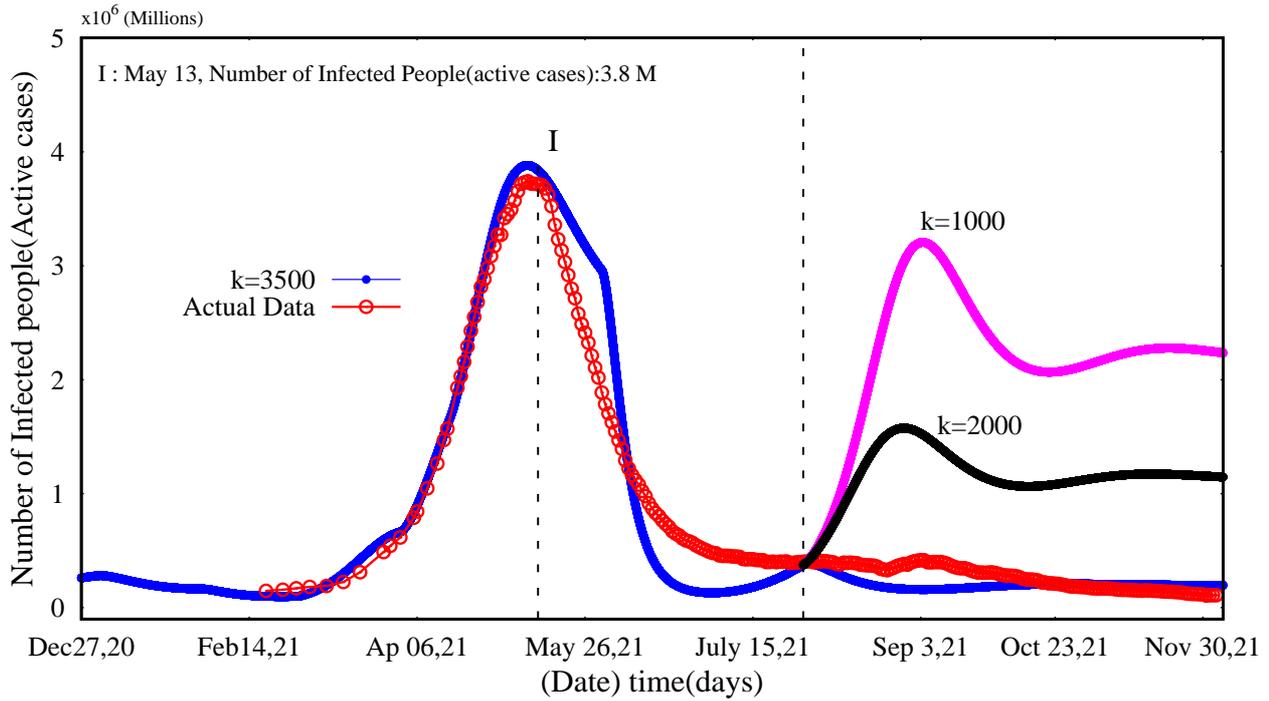}
	\caption{Numerical simulation of the number of infected individuals (after removing the number of recovered people on a particular day). The curves represent the numerical simulation of the number of infected individuals (active cases) from Dec 27, 2020, to Nov 30,  2021, with respect to the values of government action strength ($\alpha=0.2$) and different intensities of the individual reaction in the SEIR mathematical model. Data available between Dec. 2020 to July 2021 are taken for fitting the parameters. The red curve with circles corresponds to that of the actual number of infected individuals in India. The vertical black dotted lines along with label I denote the predicted maximum number of infected individuals (active cases). The pink curve and black curve show the variation of the number of infected individuals, after July 15 (second vertical lines), by considering a low value of governmental action strength ($\alpha=0.18$) and different intensities of individual reaction values ($k=1000$, pink curve and $2000$, black curve).}
\label{fig1}
\end{figure*}

\begin{figure}
	\centering
	\includegraphics[width=1.0\columnwidth]{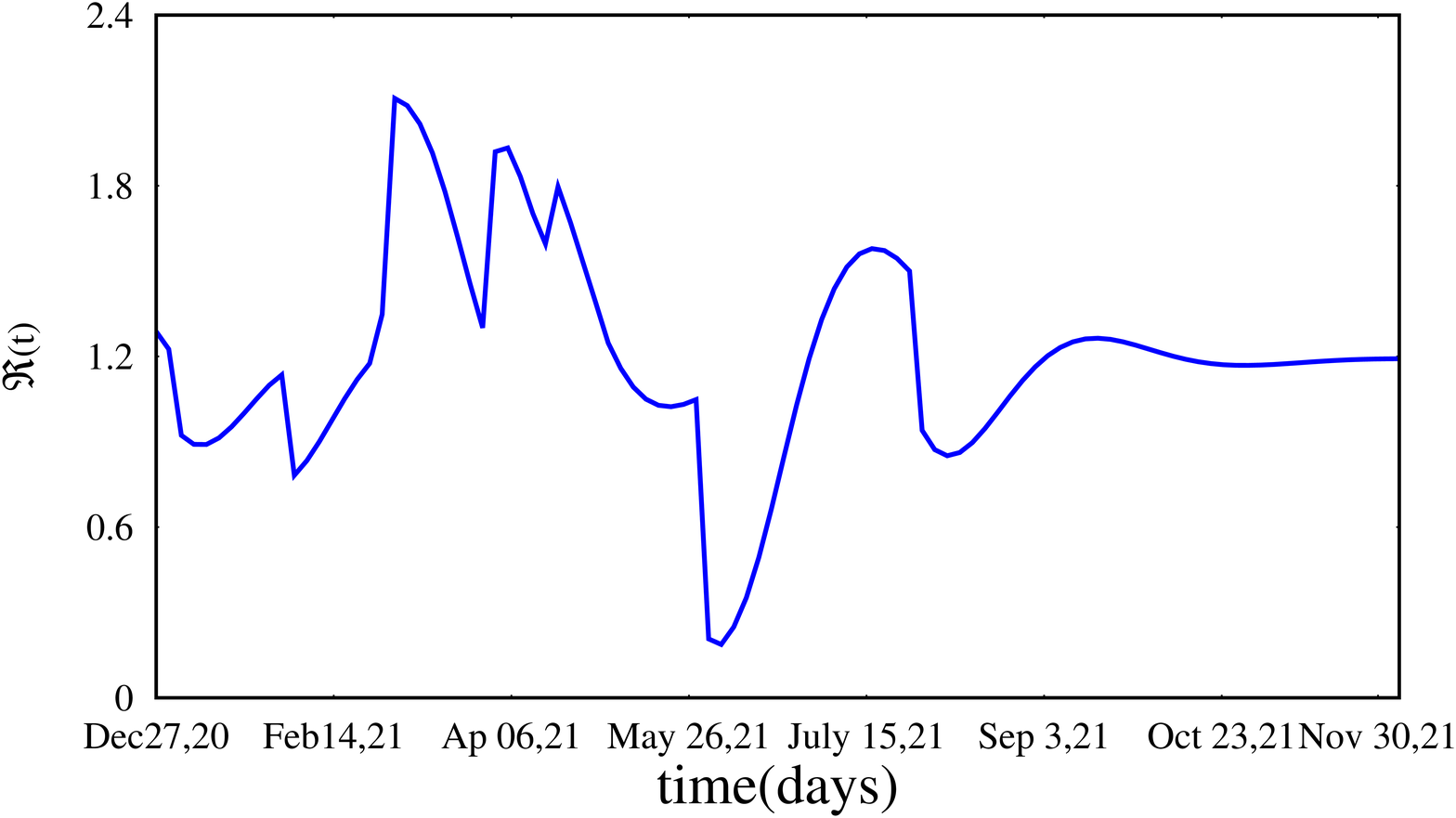}
	\caption{Variation of effective or time dependent reproduction number  $\Re(t)$ in India during the second wave of COVID-19.} 
	\label{fig2}
\end{figure}

In Ref.~\cite{gopal},  the general propagation of COVID-19 in India was reported during the period from March 25, 2020, to Dec 27, 2020,  based on the estimation of the initial value of transmission rate $\beta_{0}$ for the original outbreak of COVID-19  by considering the initial number of infected individuals in the equations for the infectious population \ref{eq1}(b)  and \ref{eq1}(c).

It is worth noting that varied transmission rates are linked to different levels of social isolation. All the parameters mentioned in (\ref{eq2}) must be varied for each location, and it depends upon the total population, the initial number of infected individuals in the whole of India, and specific states, which are necessary for the COVID-19 description. However, in order to understand the role of governmental action and intensity of individual reaction, we plotted the occurrence of the number of infected individuals for different values of the governmental action control variable $\alpha$ and the intensity of individual reaction strength $k$. For $\alpha>0$ and $k>0$, due to the certain level of governmental and individual actions to control the spread, the infected individuals tend to fall after reaching a peak value (See Fig. \ref{function}(b)). However, we note that the absence of these controls, namely $\alpha=0$ and $k=0$, can cause the infections to raise exponentially till the whole population is infected (See Fig. \ref{function}(a)) .

\begin{figure*}
	\centering
	\includegraphics[width=2.0\columnwidth]{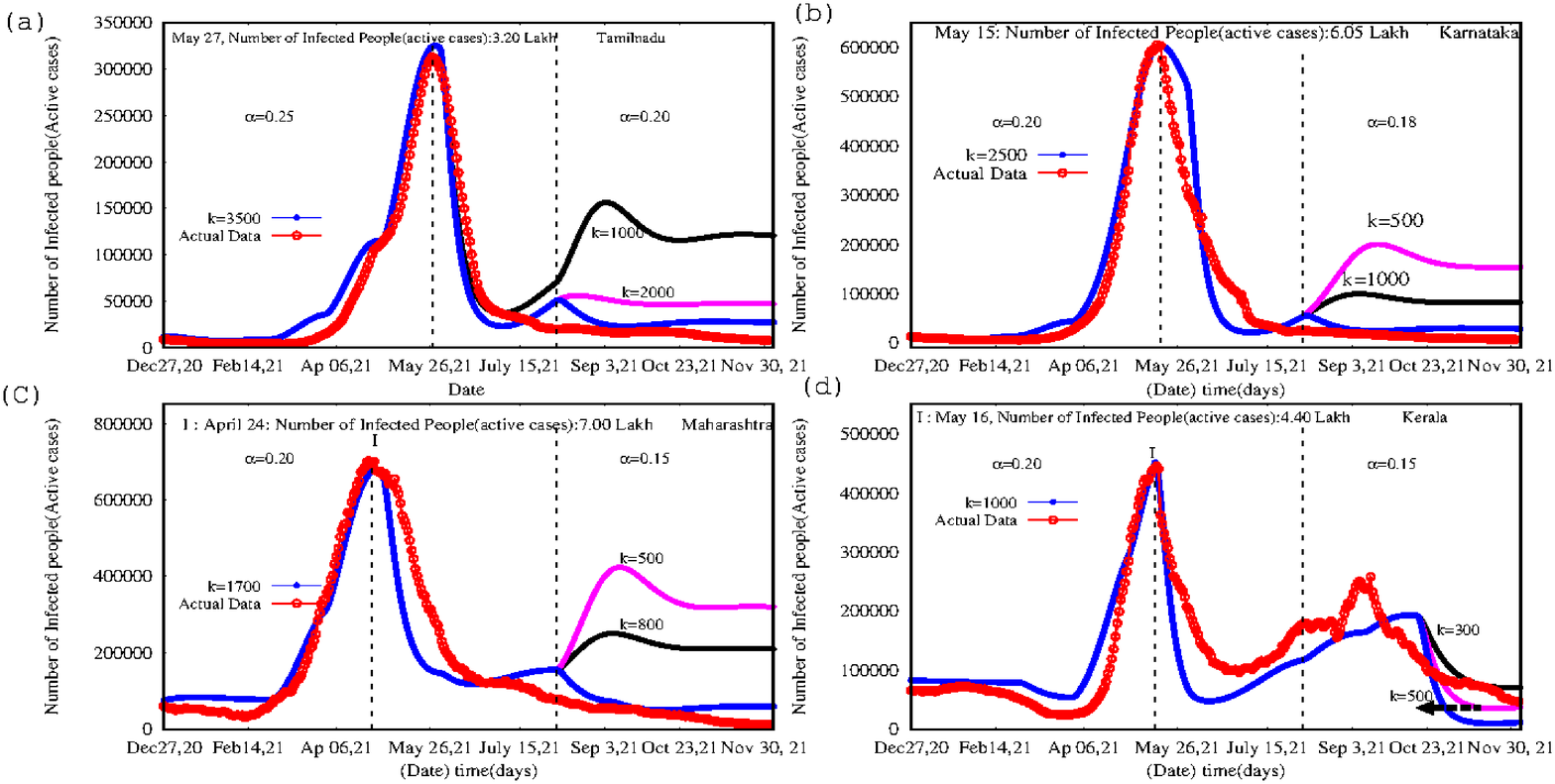}
	\caption{Numerical simulation of the number of infected individuals or active cases (after removing the number of recovered/deceased people on a particular day) in the various states of India incorporating governmental action strength $\alpha$, and different values intensity of the individual reaction $k$   : (a) Tamilnadu 
, (b) Karnataka, (c) Maharashtra  and (d) Kerala. The curves represent the numerical  simulation of the number of infected individuals (active cases) from Dec 27,2020 to Nov 30,2021 with respect to the values of government action strength and various values of intensity of the individual reaction in the SEIR mathematical model (\ref{eq1})-(\ref{eq2}).  The red curve with circles corresponds to that the actual number of infected individuals up to November 30,2021. The vertical black dotted lines along with label I denotes the predicted maximum number of infected individuals (active cases). The pink curve and black curve show the variation of the number of infected individuals, after July 15 (second vertical lines), by considering a low value of governmental action strength and different intensities of individual reaction } 
	\label{fig6}
\end{figure*}

\begin{figure*}
	\centering
	\includegraphics[width=2.0\columnwidth]{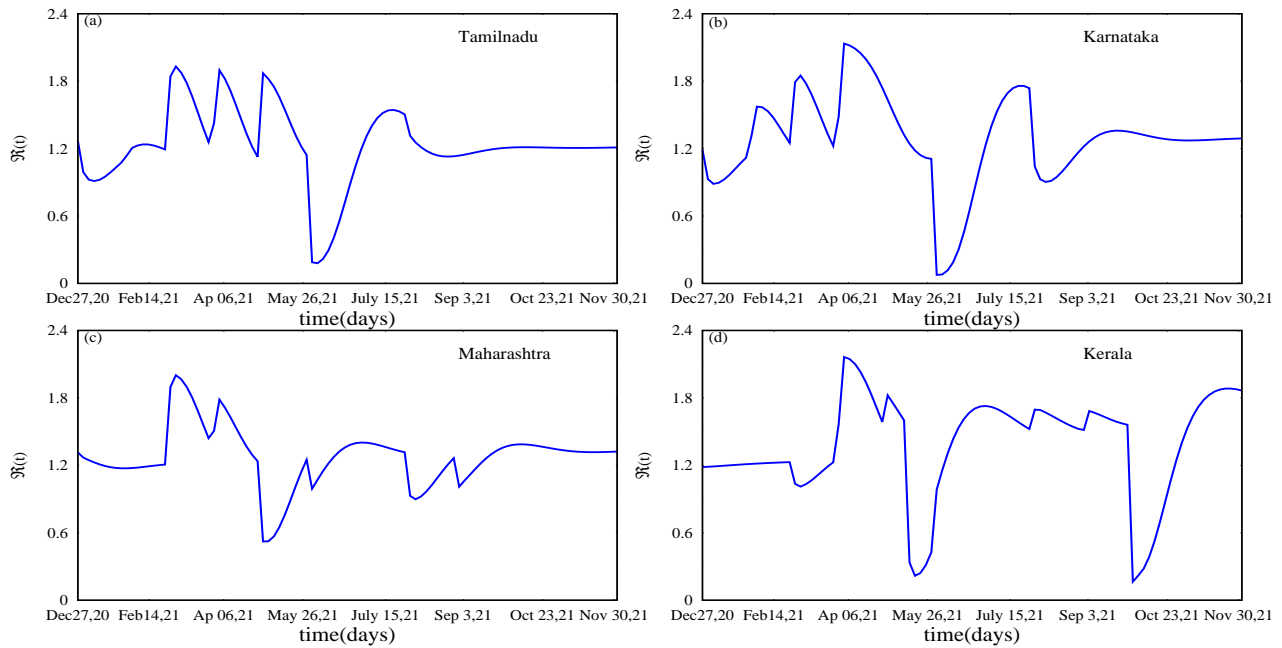}
\caption{Variation of the effective or time-dependent reproduction number $\Re(t)$ for: (a) Tamilnadu, (b) Karnataka, (c) Maharashtra  and (d) Kerala. } 
	\label{fig7}
\end{figure*}

\begin{table*}
\centering
\caption{Model parameters for Eqs.~(\ref{eq1}) and (\ref{eq2})}
\label{table1}
\begin{tabular}{lll}
\hline\noalign{\smallskip}
Parameter & Description & value/remarks/reference \\
\noalign{\smallskip}\hline\noalign{\smallskip}
$N_{0}$ & Initial number of population & India/particular state population~\cite{pop1}\\

$S_{0}$ & Initial number of susceptible population & $0.9N_{0}$ (constant)  \\

$E_{0}$ & Exposed persons for each infected person & $20I_{0}$\cite{gov1} \\
 
$I_{0}$ & Initial state of infected persons & 3 (India)/appropriate value for specific state taken from~\cite{pop1} \\
 
$\alpha$ & Government action strength  & varied in each lock-down/unlock period\\

k & intensity of individual reaction & 1117.3~\cite{lin,savi} \\

$\sigma^{-1}$ & Mean latent period & 3 (days) \\
 
$\gamma^{-1}$ & Mean infectious period & 5 (days) \\ 

$\gamma_{R}^{-1}$ & Delayed removed period & 22 (days) \\
 
$d$ & Proportion of severe cases & 0.2  \\

$\lambda^{-1}$ & Mean duration of public reaction & 11.2 (days)  \\

\noalign{\smallskip}\hline
\end{tabular}
\end{table*}

\section{Results and Discussion}

 In the present study, the daily COVID-19 active cases from the whole of India as well as specific individual states, namely Tamilnadu, Maharashtra, Kerala and Karnataka, are considered. The daily active cases are collected from the data available in~\cite{gov1}. The parameters which are presented in Table~\ref{table1} are employed for simulations with the population  $N_{0}$ and the initial number of infected individuals ($I_{0}$ which varies for each state, and the whole of India) relative to the initial outbreak in specific locations. The parameters $\alpha$ and $k$ are only needed to be adjusted for each specific period, which has become important for the estimation of the number of infected individuals. Therefore, in the following, the rate of infection based on the transmission rate (\ref{eq2}) is evaluated using the system of Eqs.(\ref{eq1}) for all the above cases.

Now, we first study the number of infected individuals in the whole of India with the strength of the individual reaction $k$ in various intervals, along with various values of governmental action strengths 
as shown in Fig.\ref{fig1}. In this case, the process of optimum selection of variables was done up to July 2021. Then, we fix constant values of $k$ and $\alpha$ in order to identify the evolution of active cases during the second wave of COVID -19. Then, we analyzed the occurrence of the number of infected individuals for different values of intensity of individual reaction strengths $k$=1000, 2000, and 3500. Here, the low and high values of $k$ represent the behavior of people along with government strategy to control the COVID -19. Therefore, the comparison of the SEIR model and real data indicates that low values of intensity of individual reaction values are not in good agreement with real data.  It shows uncontrolled growth with a large number of infected individuals for low values of $k$, and this is represented by either pink or black curve in Fig \ref{fig2} with the values of $k$ chosen as 1000 or 2000, respectively. However, the model has well fitted with a high value of the intensity of individual reaction, $k$=3500, which shows that the number of infected individuals may not increase much and it corresponds to strong measures taken by individual people and various government measures, as the epidemic curve attains low values and almost steady number of infected individuals emerge during the month October 2021 and the subsequent period.



In the case of an infectious disease like COVID -19, the number of people affected increases fast as the condition progresses, which may be measured using the time-dependent transmission rate  $\beta(t)$ in Eq.(\ref{eq2}) and it is commensurate with the effective or time-dependent reproduction number $\Re$ at time $t$ which is defined as 

\begin{align}
\Re(t)=\frac{\beta(t)}{\gamma} ,
\label{eq3}
\end{align}

The above quantity $\Re(t)$  denotes the mean number of secondary cases generated by a typical primary case at time $t$ in a population. Once a given fraction of the population has been infected and is resistant or immune, the effective reproduction number  can be used to quantify transmissibility~\cite{r2}. Because a huge part of the population will not be vulnerable to any infectious disease that causes immunity or for which a vaccine exists, an effective reproduction number can be used to estimate the severity of outbreaks in real world populations~\cite{r1}.

Further, this measure is a good identifier for the decrease or surge in infections, and also provide real-time access to the spread of the infected individuals as the value $\Re(t)>1$ indicates the growth of the newly infected cases~\cite{r1,r2,r}.  Thus the aim is to reduce the value of $\Re(t)<1$ and close to zero as much as possible. Therefore, we plotted the time-evolution of $\Re(t)$ with the help of Eqs. (\ref{eq2}-\ref{eq3}) in Fig.\ref{fig2}. It shows that the value of $\Re(t)$ is maximum during the months of April -May 2021, and has declined to near zero due to strong governmental measures,  but again increased slowly and finally settled to an almost constant value of $\Re(t)$ in the months after October 2021.

Further, analysis is also carried out for four specific states of India namely Maharashtra, Karnataka, Kerala, and Tamilnadu, using the initial population of each state $N_{0}$, and the initial number of infected individuals in the specific states (data again taken from \cite{pop1} 
and \cite{gov1}). Fig. \ref{fig6} shows the number of infected individuals from our model (\ref{eq1}) in comparison with the actual data. They clearly demonstrate the occurrence of a large number of infected individuals/active cases in each of the states during the second wave of COVID-19.  These curves also show good agreements with real data, and it mainly depends upon the individual people's reaction and adequate governmental actions in each specific state. The graphs show that the COVID-19 disease can be completely controlled if the present condition persists or if people act with greater precaution. Further, we also plotted the variation of time dependent reproduction number $\Re(t)$ with time for the above-mentioned specific states in Fig. \ref{fig7}. In Fig. \ref{fig7}, the trend of $\Re(t)$ changes over time which is different between states. One may note that  the growth rates for almost all the four states indicate that significant changes in the evolution of $\Re(t)$ during the months of April - May 2021. In Kerala, $\Re(t)$ was higher during this period, and further, maintain a larger value during the months of July - September 2021. The increasing trend in the value of $\Re(t)$ shows that there were some incidents in the particular states (partial unlocking lockdown,  people crowded in cities, religious events, etc.,)~\cite{r}.  It also shows that the value of $\Re(t)$ started to stabilize in the first week of October 2021,  implying that extensive interventions such as rigorous tracing and testing, as well as adequate clinical management of the infected individuals, specific mitigation strategies have acted effectively for each state to control the transmission COVID -19.

 Therefore, based on the above analysis, we need to realize that it is the continuous action of various state governments that is able to control this disease during the second wave of COVID 19. At the same time, it shows that the impact of the disease will be further exacerbated in the coming times if people do not take up precautionary measures i.e. like going to crowded places without wearing a mask. This is exactly what one identifies in Figure \ref{fig2} for low values of intensity of individual reaction with the low value of governmental action strength.

\begin{figure}
	\centering
	\includegraphics[width=1.0\columnwidth]{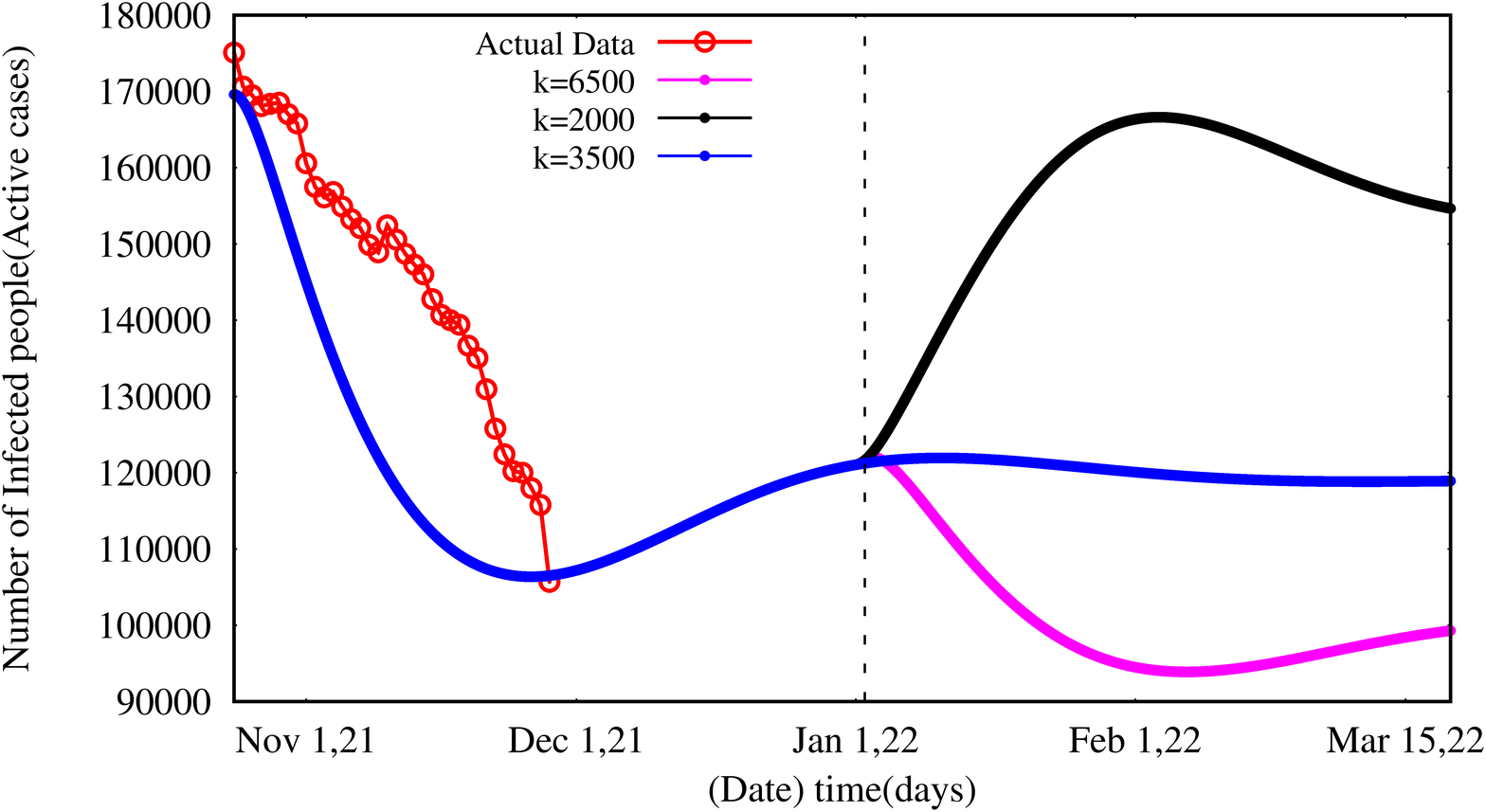}
	\caption{Numerical simulation of the number of infected individuals or active cases (after removing the number of recovered/deceased people on a particular day).  The red curve with circles corresponds to that the actual number of infected individuals up to November 30,2021. The curves represent the numerical simulation of the number of infected individuals (active cases) from November 1, 2021, to March 15,  2022, with respect to the values of the different intensities of the individual reaction in the SEIR mathematical model.  The pink curve and black curve show the variation of the number of infected individuals, after January 2022 (vertical line), by considering  different intensities of individual reaction strength. } 
	\label{third}
\end{figure}

 The various control measures implemented at the early stages during the first and second wave of COVID-19 helped the country in reducing the severity of COVID-19 spread. As the results discussed above indicate, the peak of the second wave occurred in the month of May 2021 and it is expected to gradually decline during the months of July - November 2021 (See Fig.\ref{fig1}), and it predicts around the one hundred thousand number of infected individuals at the end of November 2021~\cite{gov}. Moreover, if the second wave of COVID -19 is not fully eliminated, then there is a possibility of the third wave of COVID 19 emerging.  In order to predict the possibility of the third wave, we estimate the number of infected individuals in the month of November 2021~\cite{gov}. The results are presented in Fig. \ref{third}.
 
It shows that the third wave of COVID 19, may start after reaching a low number of infected individuals in the forthcoming period. The curves are plotted with three different values of the intensity of the individual reaction $k$, and it is clearly seen that the emergence of the third wave of COVID 19 is controllable by taking appropriate preventive measures with individual response, so that the peak will not occur (See the pink and blue curves for the values $k$=6500 and 3500, respectively). This indicates that early and timely interventions with strengthened individual reaction policies should be implemented to suppress and control the COVID-19 effectively. However, if the intensity of the individual reaction is lowered substantially, for instance when $k$=2000, the number of infected individuals may increase and in this case the curve shows a peak of COVID 19. Therefore, there is a clear indication that an increase  or steady value of the number of infected individuals will occur with respect to less or more preventive measures, respectively, after January 2022.  However, the occurrence of the number of individuals will be less when compared with the second wave of COVID 19. Further, one may note that these results depend on the available data during the second wave of COVID 19 and that we have included appropriate values of the intensity of the reaction for the control measures. 

\section{Conclusion}
 In summary, up to May, 13 2021, the total number of COVID-19 cases and the number of infected individuals (active cases) in India were 2,37,03,665 and 37,10,525, respectively and on November 30, 2021 the above counts became  3,45,87,822 and 1,00,543  respectively. Initially, the second wave of COVID-19  in India appeared as a potential threat to the country because of its rapid spread. However, the number of infected individuals got lowered in a short period only due to the public response and governmental action strengths.  Our mathematical analysis of the SEIR model appears to be an effective tool to predict the time span and to investigate epidemic evolution in India.

Our projection of the different trajectories of the pandemic, based on different scenarios (by decreasing or increasing of the value of the parameter $k$ which can change the dynamics of COVID-19 outbreak of active cases) and estimation of the effective reproduction number suggest that the second wave of COVID-19 is much more infectious than the first wave.  Further, we also examined the possible occurrence of the next wave of COVID 19 that may arise and it has been pointed out that the small or large number of infected individuals or perhaps build-up to a gentle peak, as a result of the lowering of the values of intensity of the individual reaction, can raise.

However, we find that the individual responses which may include personal hygiene, healthy habits, avoiding crowded places, wearing masks, washing hands frequently, etc. are equally or perhaps even more important than the government actions such as lock-down. The simple acts of not going out unnecessarily and keeping social distance can reduce the transmission of the virus. We have made it clear that a new outbreak of COVID-19 or a large number of infections are likely to occur if the individual action rates are reduced substantially, and that it will only begin to emerge if it is not adequately addressed by the general public, and that the onus is on the people to fully control it. Therefore, people can protect themselves against the potential third wave by adequate individual reaction: complete the vaccination process, continue to wear a masks, avoid social gatherings, continue the hand-washing, etc (which essentially refers to the high intensity of individual reaction/responses).

\section*{Acknowledgements}
The work of V.K.C. forms part of a research project sponsored by SERB-DST-MATRICS Grant No. MTR/2018/000676. M.L. wishes to thank the Department of Science and Technology for the award of a DST-SERB National Science Chair(NSC/2020/000029).

\end{document}